\documentclass[twocolumn]{article}
\usepackage[super,sort&compress,comma]{natbib} 
\usepackage{sectsty}
\usepackage{graphicx} 
\usepackage{fancyhdr}
\usepackage{lastpage}
\usepackage[format=plain,justification=justified,singlelinecheck=false,font=normal,labelfont=bf,labelsep=space]{caption} 

\begin{document}

\newcommand\blfootnote[1]{%
  \begingroup
  \renewcommand\thefootnote{}\footnote{#1}%
  \addtocounter{footnote}{-1}%
  \endgroup
}

\twocolumn[
  \begin{@twocolumnfalse}
\noindent\LARGE{\textbf{Does Rotational Melting Make Molecular Crystal Surfaces More Slippery?}}
\vspace{0.6cm}

\noindent\large{\textbf{Andrea Benassi,\textit{$^{\ast,a}$} Andrea Vanossi,\textit{$^{b,c}$}
Carlo A. Pignedoli,\textit{$^{a}$}
Daniele Passerone,\textit{$^{a}$}
 and Erio Tosatti\textit{$^{\ast,c,b,d}$}}}\vspace{0.5cm}

\noindent \normalsize{The surface of a crystal made of roughly spherical molecules exposes, above its bulk rotational phase transition at T= T$_r$, a carpet of freely rotating molecules, possibly functioning as ``nanobearings'' in sliding friction. We explored by extensive molecular dynamics simulations the frictional and adhesion changes experienced by a sliding  C$_{60}$ flake on the surface of  the prototype system C$_{60}$ fullerite. At fixed flake orientation both quantities exhibit only a modest frictional drop of order 20\% across the transition. However, adhesion and friction drop by a factor of $\sim$ 2 as the flake breaks its perfect angular alignment with the C$_{60}$ surface lattice suggesting an entropy-driven aligned-misaligned switch during pull-off at T$_r$. The results can be of relevance for sliding Kr islands, where very little frictional differences were observed at T$_r$, but also to  the sliding of C$_{60}$ -coated tip, where a remarkable factor $\sim$ 2 drop has been reported.}
\vspace{0.5cm}
 \end{@twocolumnfalse}
  ]

\blfootnote{\textit{$\ast$}~andrea.benassi@empa.ch, tosatti@sissa.it}
\blfootnote{\textit{$^{a}$~Empa-Swiss Federal Laboratories for Materials Science and Technology, CH-8600 D\"{u}bendorf, Switzerland.}}
\blfootnote{\textit{$^{b}$~ CNR-IOM Democritos National Simulation Center,Via Bonomea 265, I-34136 Trieste, Italy.}}
\blfootnote{\textit{$^{c}$~International School for Advanced Studies (SISSA), Via Bonomea 265, I-34136 Trieste, Italy.}}
\blfootnote{\textit{$^{d}$~International Centre for Theoretical Physics (ICTP), Strada Costiera 11 I-34151 Trieste, Italy}}

Exploring novel routes to achieve friction control by external physical means is a fundamental goal currently pursued
~\cite{vanossiRMP2013} in nanoscience and nanotechnology. 
The traditional lubrication control of frictional forces in macroscopic mechanical contacts is
impractical at the nanoscale, where 
contacting surfaces 
are likelier to succumb 
to capillary forces. Novel methods for control and manipulation of friction in nano and intermediate mesoscale
systems are thus constantly being explored.
As an example, mechanically induced oscillations were recently shown to reduce friction and wear, 
an effect that has been experimentally demonstrated~\cite{meyer,lantz}. 
Mismatch of relative commensurability of mutually sliding lattices may prevent 
interlocking and  stick-slip motion of the interface atoms, with a consequent friction drop (superlubricity) \cite{erdemir}.
The application of external fields (electric, magnetic, etc.) to the sliding contact was also
exploited to tune effectively the frictional response in different kind of tribological systems~\cite{liquidcristo,ionic,drummond,benassimag}.
Another, subtler route worth exploring is the possible change of adhesion and friction experienced by  a 
nanoslider when a collective property of the substrate, for example some pre-existing ordering 
is altered under the action of an external field, or of temperature.
In a given state of the substrate, its order parameter magnitude, polarization, critical fluctuations, etc.,  
determines in a unique manner the friction, affecting both the efficiency of the
slider-substrate mechanical coupling, and the rate of generation and transport of the frictional Joule
heat away from the sliding contact. A toy-model study~\cite{benassi2011} showed that the variation of 
stick-slip friction, caused by a structural order parameter switching between  
order and disorder, can indeed be large and observable. 
Hard to predict on general terms, the frictional variation occurring in a real case will
depend on the system, the mechanism, the material. 
%%%%%%%%%%%%%%%%%%%%%%%%%%%%%%%%%%%%%%%%%%%%%%%%%%%%%%%%%%%%%%%%%%%%%%%%%%%%
% Fig. 1
\begin{figure*}
\centering
\includegraphics[height=12cm]{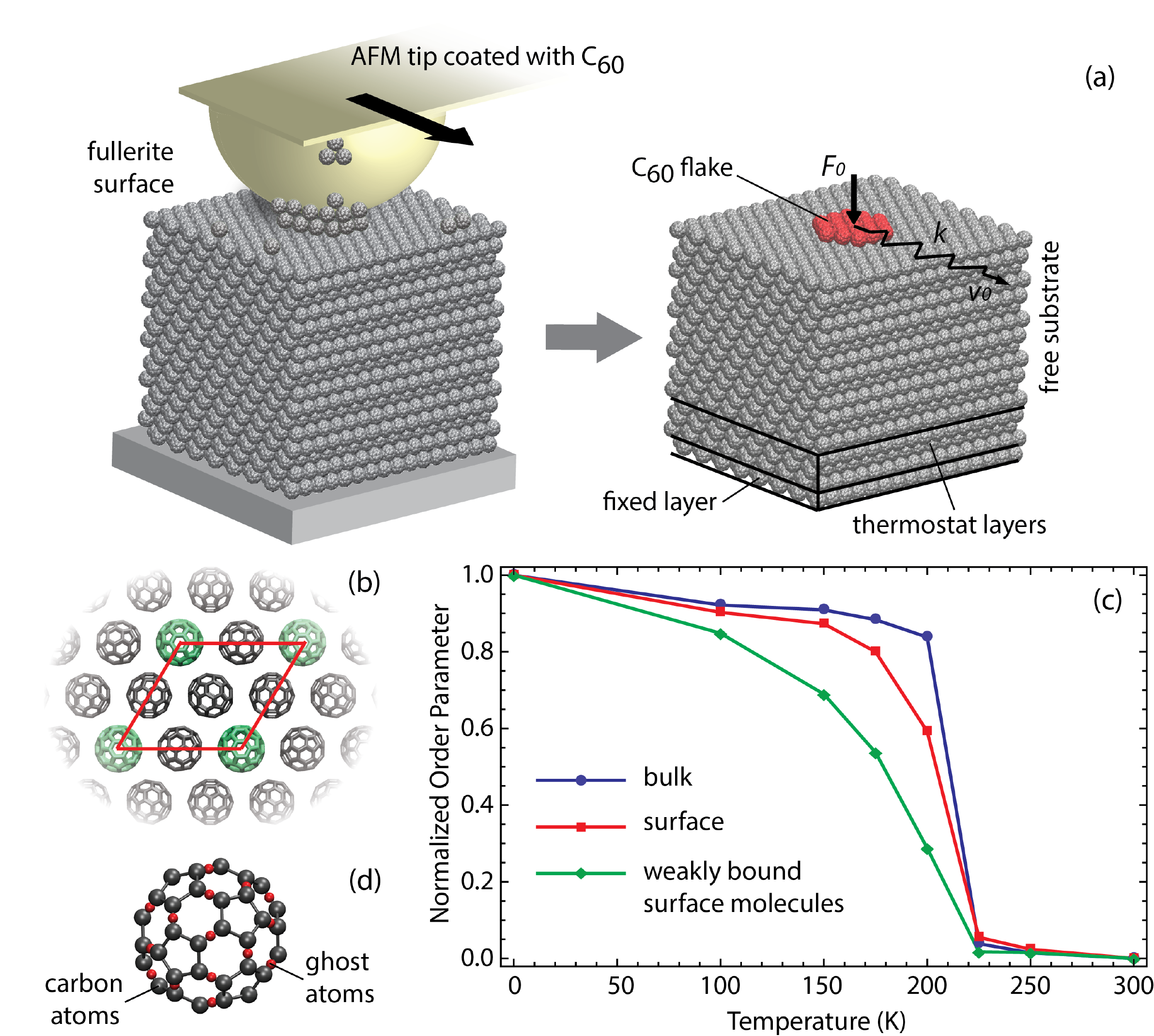}
\caption{\label{figura1} {(a) sketch of the C$_{60}$ sliding physical system (left) and of its implementation in our MD simulations (right). 
(b) Surface cell for the low temperature phase, the weakly bound surface molecules are highlighted in green. (c) variation of the rotational order parameter with temperature (see Computational Details) 
for bulk molecules (blue), all surface molecules (red) and weakly bound surface molecules (green) of panel (b). (d) interaction 
centers for the intermolecular potential of ref.~\cite{sprik}.}}
\end{figure*}
%%%%%%%%%%%%%%%%%%%%%%%%%%%%%%%%%%%%%%%%%%%%%%%%%%%%%%%%%%%%%%%%%%%%%%%%%%%%
In this work we study the friction experienced by a nanosized island or tip-attached flake sliding over 
the surface of a molecular crystal made up of nearly spherical, weakly interacting molecules, as sketched 
in figure 1 (a). We focus on the frictional effects of the rotational phase transition between
the low temperature crystalline state, positionally and rotationally ordered, and the so-called plastic 
phase at higher temperature, where molecules are still arranged in a lattice, but are rotationally
disordered. Fullerite, the insulating crystal made up of highly symmetrical  C$_{60}$ molecules, is the natural candidate
to explore this possibility. At low temperatures, the molecular centers are fcc packed and rotationally
oriented with four distinguishable molecules per cell, their axes 109$^\circ$ degrees apart (structure $Pa\bar{3}$).
The rotational order weakens upon heating, until at T=T$_r \sim$ 260 K  C$_{60}$ undergoes a first 
order plastic phase transition, with a latent heat $H \sim$ 1.2 Kcal mole$^{-1}$ corresponding to an entropy 
jump $\Delta S \sim$  2.2 k$_B$ per molecule, mainly of rotational origin. In the plastic
fcc crystal structure $Fm\bar{3}m$ the molecules rotate and are indistinguishable \cite{sachidanandam,harris}.
The bulk transition reflects directly at the C$_{60}$(111) surface, where C$_{60}$ does not evaporate and rotational 
disordering is neatly observed with different techniques~\cite{wang, goldoni}. 
The low temperature (2$ \times$2) unit cell with four inequivalent molecules, see figure 1 (b) is replaced above T$_r $ by 
(1$ \times$1) where all molecules are equivalent \cite{glebov}. Precursor surface disorder phenomena akin to early stage
surface rotational melting~\cite{tartaglino2005} also occur just below T$_r $~\cite{goldoni,benning,laforge}. 
The idea that sliding on the C$_{60}$ surface should be easier in the rotationally melted,
plastic state and the free spinning molecules could act as ``nanobearings'', has been repeatedly raised,
mostly without experimental success~\cite{braun}. In some cases the C$_{60}$ molecules may have developed a chemical 
or electrostatic interaction with the slider, thus attaching to it. When however, as in the experiment by Coffey and Krim ~\cite{coffey}, 
chemically and electrostatically inert rare gas island were observed to slide inertially over a C$_{60}$ monolayer (rotationally frozen) 
and a bilayer (rotationally melted), it was found that friction was roughly the same in both, with a decrease of 10-20\% at most, 
contrary to the large expected  ``nanobearing''  effect.  On the other hand, in alternative atomic force microscope (AFM) 
experiments it was shown that a C$_{60}$ coated  Si$_3$N$_4$ tip experienced a large drop by a factor  $\sim$ 2 of both adhesion 
and sliding friction over a C$_{60}$ surface in correspondence to the plastic transition~\cite{liang1,liang2}. 
Understanding the AFM frictional change induced by the rotational disordering, and at the same time its near absence in rare 
gas island sliding, neither effect properly understood and modeled so far, is a challenging question which we undertake here 
by computer-intensive simulation and theory, also in view of its potential interest for broader studies and applications.\\ 

Replacing the fullerene-coated tip simply by a rigid C$_{60}$ flake (n$_f$ =19 molecules), we carried out 
extensive simulations of the thermal evolution of semi-infinite fullerite approximated as a C$_{60}$ slab (n = 256
molecules/layer, N = 17 layers), thermostated at the bottom \cite{benassithermo,benassithermo2}, see figure \ref{figura1}(a).
The flake attachment/detachment, and its frictional sliding on the C$_{60}$(111) surface represent the successive equilibrium and non-equilibrium 
model problems which we analyse here highlighting the phase-transition related phenomena and their physical implications. The full
flake-surface free energies inclusive of entropic contributions are shown to depend heavily upon the relative
angular alignment between the island and the surface lattice, and are found to exhibit a detachment-induced, 
or static friction-induced,  switch from aligned to non-aligned. The switch is unfavorable at low temperature where 
stick-slip friction is high, but becomes possible at high temperature, essentially because the higher entropy favors 
the non-aligned state above the C$_{60}$ rotational transition, where friction may therefore drop.  When, by contrast,
the sliding islands are larger, superlubric, and not expected to rotate as in Kr inertial sliding, the frictional change across T$_r$ is only minor.    
%%%%%%%%%%%%%%%%%%%%%%%%%%%%%%%%%%%%%%%%%%%%%%%%%%%%%%%%%%%%%%%%%%%%%%%%%%%
% Fig.2
\begin{figure}
\centering
\includegraphics[height=12cm]{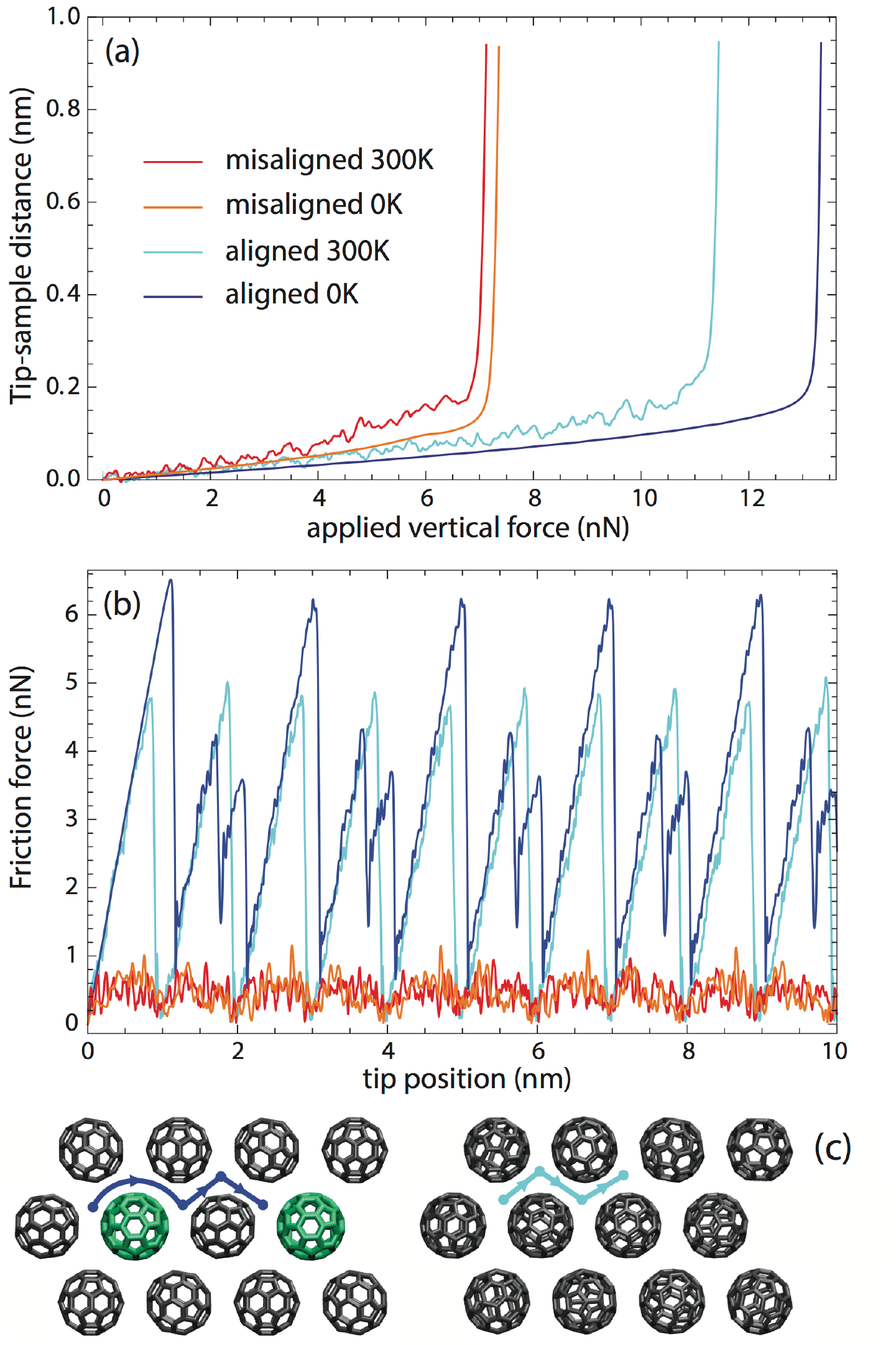}
\caption{\label{figura2} {
(a) flake-surface distance (measured relative to the ideal T= 0 aligned contact distance) as a function 
of a vertical pulling force slowly ramped up during the simulation to mimic pull-off experiments. The results are shown for two extreme temperatures, well 
below and well above the plastic transition. Note that the vertical asymptote, signaling the pull-off force, is largest for the T= 0 aligned case, and smallest 
for the T=300 K misaligned case. (b) instantaneous sliding frictional force for the flake
horizontally pulled on the surface. Bright blue, $\theta$= 0$^\circ$, T= 0 K; light blue, $\theta$= 0$^\circ$, T= 300 K; light red, $\theta$= 15$^\circ$ , 
T=0 K; bright red, $\theta$=15$^\circ$, T=300 K. With present  parameters (see Computational Details) the stick-slip sliding of the 
aligned flake is replaced by smooth sliding upon angular misalignment. (c) Trajectories of the flake center-of-mass during the stick-slip sliding in the aligned configuration at T=0 K (left) and T=300 K (right).  Circles highlight the stick sites and lines show the slip directions. At high temperature surface molecules are indistinguishable and the flake proceeds by zig-zag through the hollow sites. At low temperatures the moving flake preferentially promotes the rotation of the weakly bound BS molecule (green) and jumps forward, with energy release and higher friction.}} 
\end{figure}
%
%%%%%%%%%%%%%%%%%%%%%%%%%%%%%%%%%%%%%%%%%%%%%%%%%%%%%%%%%%%%%%%%%%%%%%%%%%%%
We adopt the C$_{60}$ - C$_{60}$ distance and angle- dependent interactions of Sprik et al.~\cite{sprik}, 
which yield the bulk plastic phase transition temperature T$_{r} \sim$ 230 K, latent heat 
$ H_m \sim$ 1.16 Kcal mole$^{-1}$, and entropy jump $\Delta S\sim$ 1.5 k$_B$ mole$^{-1}$, reproducing very reasonably 
the experimental values.
The C$_{60}$ flake, initially brought in optimal contact with C$_{60}$(111), was driven in two different ways. First, to measure the adhesive force,
it was lifted off vertically with adiabatically slow speed and in thermal equilibrium, from close contact out to large distance. Second, to measure the
sliding friction, the flake was dragged with a speed $v_0$ of 1- 10 m/s parallel to the surface through a spring 
of constant  $k \sim 7$ N m$^{-1}$, chosen in such a way that the sliding could occur by stick-slip as it does in real low temperature AFM , of course
at much lower speed. 
Especially on account of the long-range forces the simulation size was still very large and time-consuming, and meaningfully
fixed flake orientations had to be carefully picked.  We chose two orientation angles of the flake lattice relative to the crystal surface,
namely zero (aligned, commensurate flake) and 15$^\circ$ (misaligned, nearly incommensurate flake). 
Simulations covered a temperature grid from zero to 300 K, and meaningful averages for each point of the grid were collected. 
Typical force results of the frictional simulations are displayed in figure ~\ref{figura2}(a) and (b), where
the modest smooth friction of the misaligned flake strongly contrasts with the large stick-slip friction of the aligned flake. 
The center-of-mass stick-slip trajectory interestingly shows a zig-zag pattern which, as indicated in figure 2, is symmetrical 
on the high temperature (1$ \times$1) surface, but turns asymmetrical below T$_r$ where one out of four molecules
in the (2$ \times$2) cell (dubbed BS for "black sheep" in Ref. ~\cite{laforge}) is inequivalent to the remaining three. 
The flake detachment simulations were carried out with slow, nearly adiabatic speed, monitoring force and total potential
energy at all steps. That also permitted the  free energy of detachment (obtained by force integration) and the internal  
energy of detachment to be compared, whereby the (negative) entropy of adhesion could be monitored as a function of
distance.\\

The resulting force of total pull-off and the average sliding frictional force of the flake are shown in figure 3.  For each of the two fixed flake orientations 
both the pull-off and friction decrease gently upon heating. At the transition temperature T$_{r}$  they show a drop
as anticipated, however a relatively modest one, no more than about 10--20\%  
(see the drop in the blue curves across the transition). The relative frictional drop does not increase 
if we allow the C$_{60}$ flake molecules to rotate instead of being rotationally rigid. We thus conclude that
there is no simple nanobearing effect, in agreement with Kr sliding islands by Coffey and Krim~\cite{coffey}.  
%%%%%%%%%%%%%%%%%%%%%%%%%%%%%%%%%%%%%%%%%%%%%%%%%%%%%%%%%%%%%%%%%%%%%%%%%%%%%
% Fig.3
\begin{figure}
\centering
\includegraphics[height=10cm]{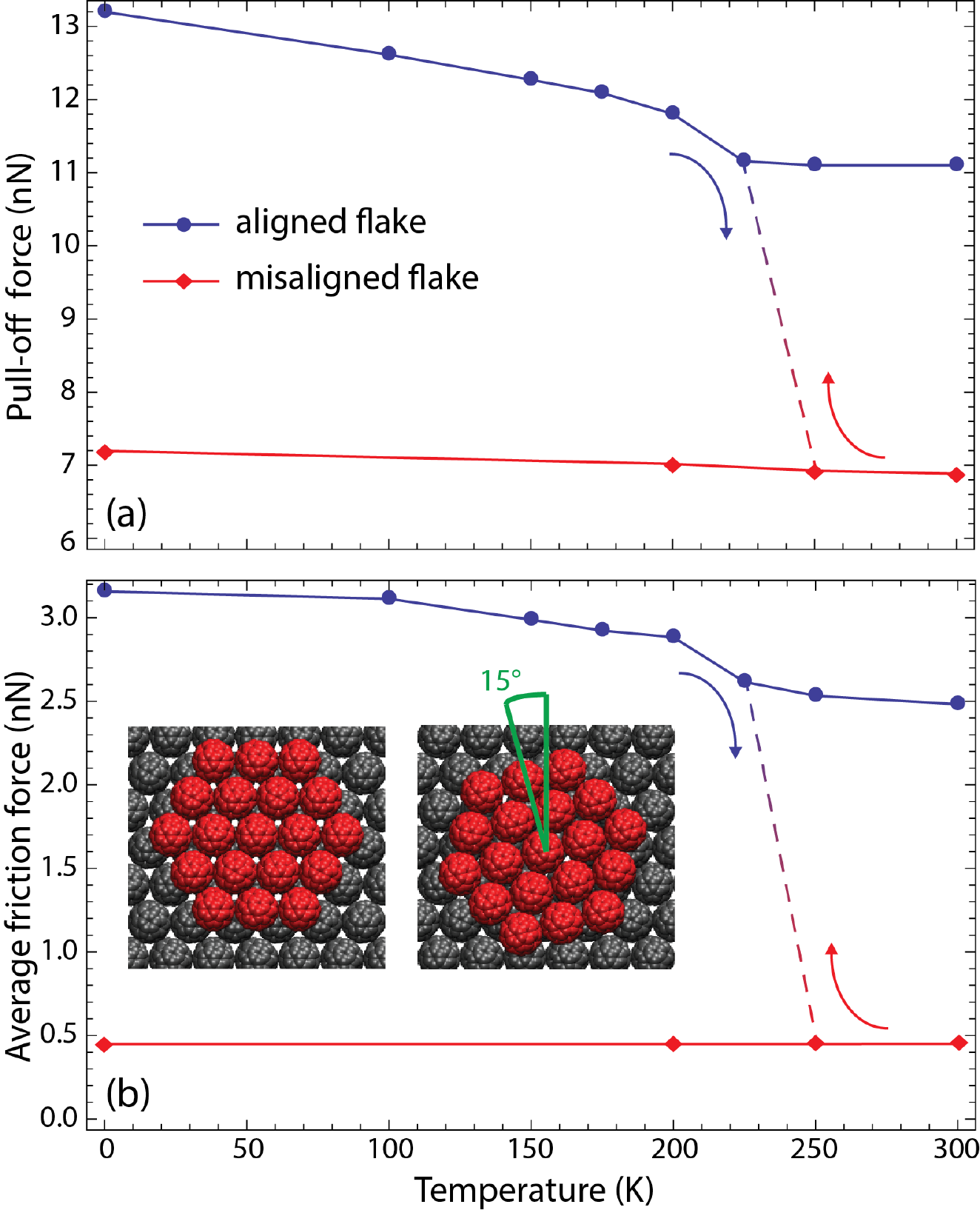}
\caption{\label{figura3} {(a) Simulated pull-off force and (b) average sliding friction force of a C$_{60}$ flake on a C$_{60}$ substrate as a function of temperature. Notice the average thermal decrease of both quantities, and the moderate $\simeq$ 10--20\%  drop at the rotational disordering (plastic) transition. Error bars are slightly smaller than the dot size.
}}
\end{figure}
%
%%%%%%%%%%%%%%%%%%%%%%%%%%%%%%%%%%%%%%%%%%%%%%%%%%%%%%%%%%%%%%%%%%%%%%%%%%%%%
However, the experimentally reported drop of friction and adhesion is an order of magnitude larger, namely about a factor 2, for a C$_{60}$-covered AFM tip~\cite{liang1,liang2}. 
On the other hand our simulations show that adhesion and friction of the misaligned flake are about a factor 2--3 smaller than those of the 
aligned  flake, owing to the much worse flake-surface interlocking caused by misalignment (compare red and blue curves). 
We are thus led to suspect a switch from aligned (commensurate) to misaligned (incommensurate) taking place in the AFM experiments
at the plastic transition temperature T$_{r}$ of the C$_{60}$ substrate, a hypothesis that could explain the different outcome 
of the Kr sliding and of the AFM tip experiment. Not nearly enough is known about the tip and the details of its  C$_{60}$ 
coating to explain how this could happen.  However we have found that we can use our simulation forces to build a free energy 
scenario that is quite suggestive.\\

We start by noting that adhesion and friction behave extremely similarly and literally go hand in hand, even dropping by a very
similar factor at the transition, both in our simulations and in the AFM experiments. Thus we can concentrate on adhesion, 
which is an equilibrium property and as such simpler to extract.  
For fixed angular alignment, namely, $\theta$ =0 or $\theta$ = 15 $^{\circ}$, we obtain the flake-surface  free energy 
in full thermal equilibrium as a function of distance by integrating the vertical (negative) pulling force $f(z)$ (shown on figure 4)
\begin{equation}
F(z)=\int_{\infty}^{z} f(z') dz'
\end{equation}
If $z_0 = z_0(\theta, T) $ is the equilibrium, force free vertical coordinate of the flake-surface contact, where $f(z_0)$ = 0,
the adhesion free energy is just $W = - F(z_0)$.  
The calculated free energies are plotted in figure 4 for aligned and misaligned flakes for the two limiting cases of $T=0$ and $T=300$ K.

As expected, adhesion is stronger for the commensurate flake, both below and above T$_{r}$.
The flake-surface equilibrium contact distance is larger by about 0.75--1 $\mathrm {\AA}$ in the misaligned case, due to ill fitting
of the flake into the  C$_{60}$ surface lattice. Owing to that, the misaligned free energies actually fall {\it below} the aligned 
free energy beyond some distance, suggesting a  possible switching mechanism from aligned to misaligned during detachment.
The same switch can then occur during stick-slip sliding, since as suggested by simulation the slip-related 
relief of strain accumulated during sticking pushes the flake outwards, similar  to detachment.  The close parallelism 
of pull-off and frictional forces is experimentally solid.\\

At low temperatures the fixed angle aligned-misaligned energies permit no switch until a very large detachment distance, 
consistent with adhesion, pull-off and friction forces that are large, in turn agreeing with experiment~\cite{liang1,liang2}.
At higher temperature above T$_r$, the situation may change and the aligned-misaligned switching can take  place
at much closer distances, possibly already close to contact, because of entropic reasons.  That conclusion is not immediately
suggested by the fixed-angle misaligned free energy results of figure 4, where $\theta$ =0 and $\theta$ = 15$^{\circ}$
only cross at large flake-surface separation. However, $\theta$ =0 is an isolated  low energy state of the flake and its free energy
is well described by fixed- $\theta$ simulations, but the same is not true for the fixed-angle misaligned state $\theta$ = 15$^{\circ}$. 
In fact, the $\theta \neq  0$  states are a full continuum of states of very similar energy 
spanning all angles from zero to 2$\pi$, and of all (x,y) positions as permitted by the ``superlubric'' nature of 
a misaligned flake, a situation conceptually similar to that of a graphene flake 
on graphite \cite{dienwiebel,tt} but with molecular rotations added on top.  The free molecular rotations 
add  a larger amount of entropy to the misaligned states because they further multiply the number of configurations 
made accessible by the facile flake rotations and translations. The continuum of non-aligned states may  thus
become entropically favored causing the aligned-misaligned switch precisely at the rotational phase transition point. 
Although this entropy driven switch should in principle be amenable to direct verification in our model,  extreme computational 
weight makes it prohibitive to carry out the simulation 
effort necessary to calculate the extra entropy and consequent free energy gain of the manifold of misaligned 
and superlubric states. Thus we cannot yet prove that they will indeed prevail  above T$_r$ relative to the energetically 
preferred aligned and pinned state below $T_r$. Despite this hurdle, this scenario is rendered highly probable 
by our fixed-angle partial results already, suggesting as in figure 3 that the occurrence of the switch at T$_r$ neatly 
and even quantitatively explains the jump of adhesion and friction observed by AFM.     
As further indicated by the red dotted line speculatively added in figure 4 the freedom associated with 
flake rotations and translations will give rise, above T$_r$, to a large extra 
negative entropic contribution to all configurations {\it except} the aligned flake at close contact, which is then
effectively raised favoring a much earlier switching to a misaligned, low adhesion, low  friction state above T$_r$.
%%%%%%%%%%%%%%%%%%%%%%%%%%%%%%%%%%%%%%%%%%%%%%%%%%%%%%%%%%%%%%%%%%%%%%%%%%%%%%
% Fig.4
\begin{figure}
\centering
\includegraphics[height=10cm]{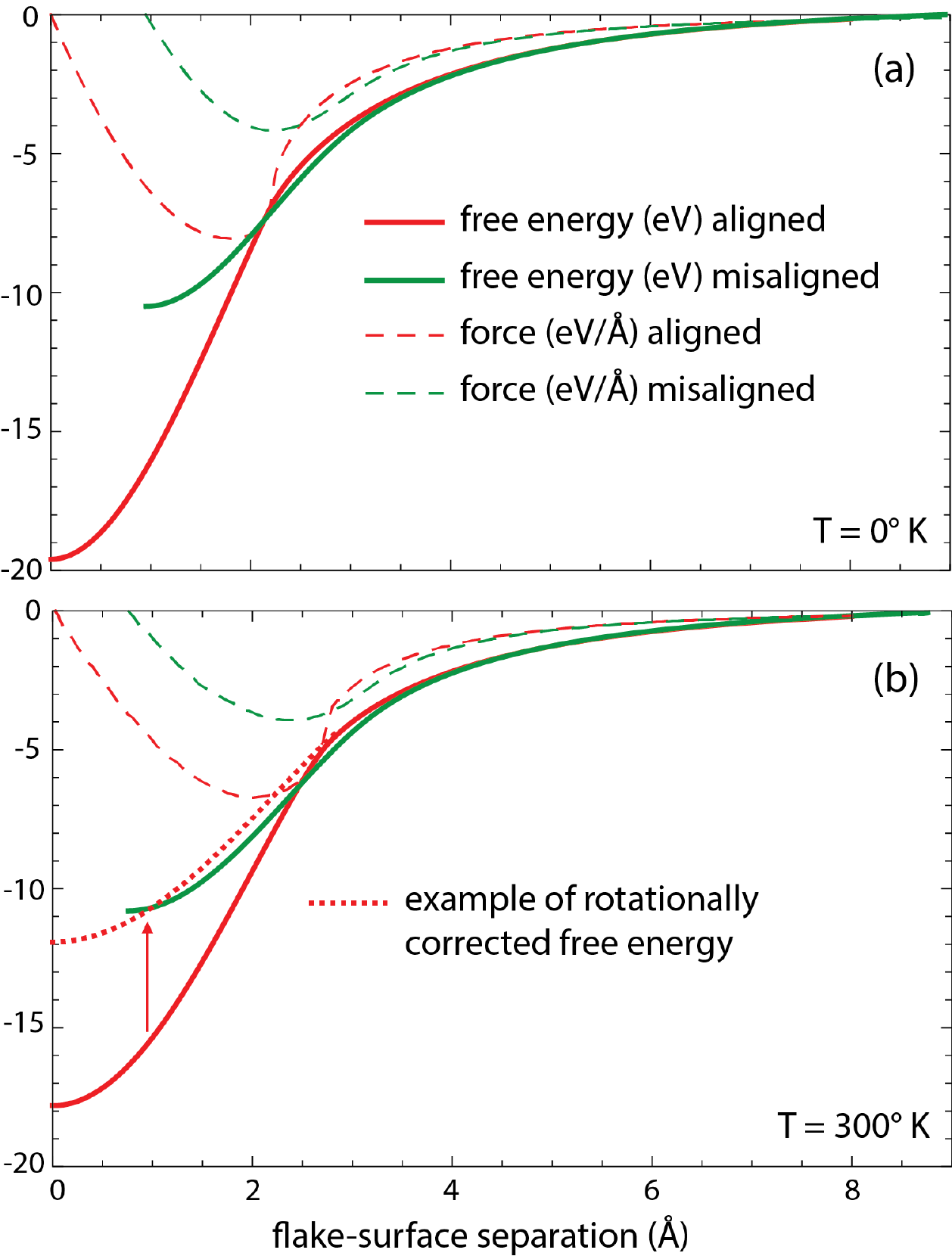}
\caption{\label{figura4}{Free energy comparison of the aligned ($\theta$ =0$^{\circ}$) and of the 
misaligned  ($\theta$ =15$^{\circ}$) states of a C$_{60}$ flake on the C$_{60}$ surface lattice (solid lines) and corresponding attractive forces versus 
detachment distance $z$ at (a) T= 0 K and (b) T=300 K (dashed lines). The aligned-misaligned free energy crossing suggests 
a switching between aligned and misaligned taking place upon detachment. Whereas in the fixed angle approximation used in these simulations the 
crossing only occurs at large distance, the aligned state lacks the large additional entropy of the
misaligned or of the detached state, an effect not included in the angle-rigid simulation. The qualitative effect of that extra entropy may  give rise as 
indicated by the speculative red dotted line to a much readier aligned-misaligned switch above T$_r$, explaining the AFM friction and adhesion 
jump indicated in figure 3}}
\end{figure}
%
%%%%%%%%%%%%%%%%%%%%%%%%%%%%%%%%%%%%%%%%%%%%%%%%%%%%%%%%%%%%%%%%%%%%%%%%%%%%%%%
We reiterate here that the same difference between low and high temperatures does not apply to the case of inertially sliding Kr on 
C$_{60}$, where Kr islands are of mesoscopically large size and have very small grip on the C$_{60}$ surface,  
and where the data of Coffey and Krim indeed indicate free incommensurate sliding for either rotating or frozen C$_{60}$
with a small slip time difference well compatible with our 10--20\%  result.\\   

In conclusion, theory supports and explains the negative result of Coffey and Krim in their search for rolling lubricity over 
rotationally disordered C$_{60}$. It also leads to seriously consider the physically interesting possibility that entropy 
could be at the origin of the observed AFM frictional and adhesion downwards jump observed over the C$_{60}$ surface where a switch is  suggested from an 
angularly arrested, pinned and aligned state of the contact flake below T$_r$, to a freely rotating and translating manifold of misaligned 
states permitted by the rotating C$_{60}$ molecules -- an appealing  ``nanobearing'' effect, directly caused by a substrate
phase transition.

\section*{Computational Details}
The simulations have been performed using the LAMMPS code~\cite{lammps}. Every C$_{60}$ molecule is treated as a rigid body in the quaternion representation.
The time integration is based on the Richardson iterative (approximate) method and thus is not fully energy conserving. This does not constitute a problem as long as one works in the canonical ensemble using stochastic thermostats.\\
The simulation of a large portion of the fullerite substrate of 17 layers 16$\times$16 molecules wide (4352 molecules) is necessary both to reproduce with good accuracy the 
phase transition and to favour the thermalization of the energy released by the sliding tip.
The substrate bottom layer is fixed by blocking the translational motion of each molecule but preserving its rotational degrees of freedom, a complete freezing of all the molecule 
degrees of freedom might, in fact, hinder the rotational melting transition.\\
A Langevin thermostat is applied to the 3 layers immediately above the fixed one, it is applied directly to the molecule atoms thus thermostatating both the rotational and vibrational 
degrees of freedom. The thermostat guarantees the onset of a steady state condition absorbing the Joule heat constantly introduced in the simulation cell by the external forces acting 
on the flake. At the same time, positioning the thermostat far away form the surface, one prevents the viscous damping term from affecting spuriously the sliding 
dynamics. The value of the viscous damping coefficient has been chosen according to the calibration procedure described in ref.~\cite{benassithermo,benassithermo2} in order to 
maximize the energy absorption rate minimizing phonon reflection from the cell boundaries.\\
According to the Sprik potential~\cite{sprik} every molecule contains 60 interaction centers corresponding to the carbon atoms plus 30 interaction centers located on the double bonds 
to mimic the charge delocalization there occurring, see figure 1 (d), thus the simulation cell contains roughly 394000 particles. 
Every interaction center has short range (Van der Waals) and long range (Coulomb) contributions, the latter is treated in reciprocal 
space with the particle-particle-particle-mesh method~\cite{pppm}.\\
The phase transition has been characterized calculating the order parameter~\cite{laforge}: 
\begin{equation}
S(T)=\frac{1}{N*60}\sum_{i=1}^{N}\sum_{j=1}^{60}  S_{6,1}(\theta_{i,j},\phi_{i,j})
\end{equation}
where $i$ runs over the $N$ molecules of the simulation cell, $j$ runs over the 60 carbon atoms of every molecule, $S_{6,1}(\theta,\phi)$ is the icosahedral \emph{symmetry adapted function}, i.e. a proper combination of spherical harmonics giving rise to a function with the same symmetry of the $C_{60}$ molecule. For every atom, the angles $\theta$ and $\phi$ are defined with respect to a molecular reference frame with $z$-axis lying on a 5-fold symmetry axis and $y$-axis lying on a 2-fold symmetry axis. The order parameter is normalized dividing it by its zero temperature value $S_{6,1}(\theta,\phi)(T)/S_{6,1}(\theta,\phi)(0)$.\\

\section*{Acknowledgements}
This work was supported by grant CRSII2 136287/1 of the Swiss National Science Foundation and it is also part of  ERC Advanced Grant No. 320796-
MODPHYSFRICT. This work is also supported in part by COST Action MP1303. Computational resources for the large scale MD simulations have been 
provided by the PRACE project 2012071262. The authors are grateful to Andrew Jewett from University of California (Santa Barbara) for the technical 
support with the LAMMPS code.\\ 

%\footnotesize{
\bibliography{biblio}

\providecommand*{\mcitethebibliography}{\thebibliography}
\csname @ifundefined\endcsname{endmcitethebibliography}
{\let\endmcitethebibliography\endthebibliography}{}
\begin{mcitethebibliography}{28}
\providecommand*{\natexlab}[1]{#1}
\providecommand*{\mciteSetBstSublistMode}[1]{}
\providecommand*{\mciteSetBstMaxWidthForm}[2]{}
\providecommand*{\mciteBstWouldAddEndPuncttrue}
  {\def\EndOfBibitem{\unskip.}}
\providecommand*{\mciteBstWouldAddEndPunctfalse}
  {\let\EndOfBibitem\relax}
\providecommand*{\mciteSetBstMidEndSepPunct}[3]{}
\providecommand*{\mciteSetBstSublistLabelBeginEnd}[3]{}
\providecommand*{\EndOfBibitem}{}
\mciteSetBstSublistMode{f}
\mciteSetBstMaxWidthForm{subitem}
{(\emph{\alph{mcitesubitemcount}})}
\mciteSetBstSublistLabelBeginEnd{\mcitemaxwidthsubitemform\space}
{\relax}{\relax}

\bibitem[Vanossi \emph{et~al.}(2013)Vanossi, Manini, Urbakh, Zapperi, and
  Tosatti]{vanossiRMP2013}
A.~Vanossi, N.~Manini, M.~Urbakh, S.~Zapperi and E.~Tosatti, \emph{Rev. Mod.
  Phys.}, 2013, \textbf{85}, 529\relax
\mciteBstWouldAddEndPuncttrue
\mciteSetBstMidEndSepPunct{\mcitedefaultmidpunct}
{\mcitedefaultendpunct}{\mcitedefaultseppunct}\relax
\EndOfBibitem
\bibitem[Socoliuc \emph{et~al.}(2006)Socoliuc, Gnecco, Maier, Pfeiffer,
  Baratoff, Bennewitz, and E.Meyer]{meyer}
A.~Socoliuc, E.~Gnecco, S.~Maier, O.~Pfeiffer, A.~Baratoff, R.~Bennewitz and
  E.Meyer, \emph{Science}, 2006, \textbf{313}, 207\relax
\mciteBstWouldAddEndPuncttrue
\mciteSetBstMidEndSepPunct{\mcitedefaultmidpunct}
{\mcitedefaultendpunct}{\mcitedefaultseppunct}\relax
\EndOfBibitem
\bibitem[Lantz \emph{et~al.}(2009)Lantz, Wiesmann, and Gotsmann]{lantz}
M.~A. Lantz, D.~Wiesmann and B.~Gotsmann, \emph{Nat. Nanotech.}, 2009,
  \textbf{4}, 586\relax
\mciteBstWouldAddEndPuncttrue
\mciteSetBstMidEndSepPunct{\mcitedefaultmidpunct}
{\mcitedefaultendpunct}{\mcitedefaultseppunct}\relax
\EndOfBibitem
\bibitem[Erdemir and Martin(2007)]{erdemir}
A.~Erdemir and M.~Martin, \emph{Superlubricity}, Elsevier, New York, USA,
  2007\relax
\mciteBstWouldAddEndPuncttrue
\mciteSetBstMidEndSepPunct{\mcitedefaultmidpunct}
{\mcitedefaultendpunct}{\mcitedefaultseppunct}\relax
\EndOfBibitem
\bibitem[Kimura \emph{et~al.}(1994)Kimura, Nakano, Kato, and
  Morishita]{liquidcristo}
Y.~Kimura, K.~Nakano, T.~Kato and S.~Morishita, \emph{Wear}, 1994,
  \textbf{175}, 143\relax
\mciteBstWouldAddEndPuncttrue
\mciteSetBstMidEndSepPunct{\mcitedefaultmidpunct}
{\mcitedefaultendpunct}{\mcitedefaultseppunct}\relax
\EndOfBibitem
\bibitem[Sweeney \emph{et~al.}(1012)Sweeney, Hausen, Hayes, Webber, Endres,
  Rutland, Bennewitz, and Atkin]{ionic}
J.~Sweeney, F.~Hausen, R.~Hayes, G.~B. Webber, F.~Endres, M.~W. Rutland,
  R.~Bennewitz and R.~Atkin, \emph{Phys. Rev. Lett.}, 1012, \textbf{109},
  155502\relax
\mciteBstWouldAddEndPuncttrue
\mciteSetBstMidEndSepPunct{\mcitedefaultmidpunct}
{\mcitedefaultendpunct}{\mcitedefaultseppunct}\relax
\EndOfBibitem
\bibitem[Drummond(2012)]{drummond}
C.~Drummond, \emph{Phys. Rev. Lett.}, 2012, \textbf{109}, 154302\relax
\mciteBstWouldAddEndPuncttrue
\mciteSetBstMidEndSepPunct{\mcitedefaultmidpunct}
{\mcitedefaultendpunct}{\mcitedefaultseppunct}\relax
\EndOfBibitem
\bibitem[Benassi \emph{et~al.}(2014)Benassi, Schwenk, Marioni, Hug, and
  Passerone]{benassimag}
A.~Benassi, J.~Schwenk, M.~A. Marioni, H.~J. Hug and D.~Passerone,
  \emph{Advanced Materials Interfaces}, 2014, \textbf{1}, 1400023\relax
\mciteBstWouldAddEndPuncttrue
\mciteSetBstMidEndSepPunct{\mcitedefaultmidpunct}
{\mcitedefaultendpunct}{\mcitedefaultseppunct}\relax
\EndOfBibitem
\bibitem[Benassi \emph{et~al.}(2011)Benassi, Vanossi, Santoro, and
  Tosatti]{benassi2011}
A.~Benassi, A.~Vanossi, G.~Santoro and E.~Tosatti, \emph{Rev. Rev. Lett.},
  2011, \textbf{106}, 256102\relax
\mciteBstWouldAddEndPuncttrue
\mciteSetBstMidEndSepPunct{\mcitedefaultmidpunct}
{\mcitedefaultendpunct}{\mcitedefaultseppunct}\relax
\EndOfBibitem
\bibitem[Sprik \emph{et~al.}(1992)Sprik, Cheng, and Klein]{sprik}
M.~Sprik, A.~Cheng and M.~Klein, \emph{J. Phys. Chem.}, 1992, \textbf{96},
  2027\relax
\mciteBstWouldAddEndPuncttrue
\mciteSetBstMidEndSepPunct{\mcitedefaultmidpunct}
{\mcitedefaultendpunct}{\mcitedefaultseppunct}\relax
\EndOfBibitem
\bibitem[Sachidanandam and Harris(1991)]{sachidanandam}
R.~Sachidanandam and A.~B. Harris, \emph{Phys. Rev. Lett.}, 1991, \textbf{67},
  1467\relax
\mciteBstWouldAddEndPuncttrue
\mciteSetBstMidEndSepPunct{\mcitedefaultmidpunct}
{\mcitedefaultendpunct}{\mcitedefaultseppunct}\relax
\EndOfBibitem
\bibitem[Harris and Sachidanandam(1992)]{harris}
A.~B. Harris and R.~Sachidanandam, \emph{Phys. Rev. B}, 1992, \textbf{46},
  4944\relax
\mciteBstWouldAddEndPuncttrue
\mciteSetBstMidEndSepPunct{\mcitedefaultmidpunct}
{\mcitedefaultendpunct}{\mcitedefaultseppunct}\relax
\EndOfBibitem
\bibitem[Wang \emph{et~al.}(2001)Wang, Zeng, Wang, Hou, Li, and Yang]{wang}
H.~Wang, C.~Zeng, B.~Wang, J.~G. Hou, Q.~Li and J.~Yang, \emph{Phys. Rev. B},
  2001, \textbf{63}, 085417\relax
\mciteBstWouldAddEndPuncttrue
\mciteSetBstMidEndSepPunct{\mcitedefaultmidpunct}
{\mcitedefaultendpunct}{\mcitedefaultseppunct}\relax
\EndOfBibitem
\bibitem[Goldoni \emph{et~al.}(1996)Goldoni, Cepek, and Modesti]{goldoni}
A.~Goldoni, C.~Cepek and S.~Modesti, \emph{Phys. Rev. B}, 1996, \textbf{54},
  2890\relax
\mciteBstWouldAddEndPuncttrue
\mciteSetBstMidEndSepPunct{\mcitedefaultmidpunct}
{\mcitedefaultendpunct}{\mcitedefaultseppunct}\relax
\EndOfBibitem
\bibitem[Glebov \emph{et~al.}(1997)Glebov, Senz, Toennies, and
  Gensterblum]{glebov}
A.~Glebov, V.~Senz, J.~Toennies and G.~Gensterblum, \emph{J. Appl. Phys.},
  1997, \textbf{82}, 2329\relax
\mciteBstWouldAddEndPuncttrue
\mciteSetBstMidEndSepPunct{\mcitedefaultmidpunct}
{\mcitedefaultendpunct}{\mcitedefaultseppunct}\relax
\EndOfBibitem
\bibitem[Tartaglino \emph{et~al.}(2005)Tartaglino, Zykova-Timan, Ercolessi, and
  Tosatti]{tartaglino2005}
U.~Tartaglino, T.~Zykova-Timan, F.~Ercolessi and E.~Tosatti, \emph{Physics
  Reports}, 2005, \textbf{411}, 291\relax
\mciteBstWouldAddEndPuncttrue
\mciteSetBstMidEndSepPunct{\mcitedefaultmidpunct}
{\mcitedefaultendpunct}{\mcitedefaultseppunct}\relax
\EndOfBibitem
\bibitem[Benning \emph{et~al.}(1993)Benning, Stepniak, and Weaver]{benning}
P.~J. Benning, F.~Stepniak and J.~H. Weaver, \emph{Phys. Rev. B}, 1993,
  \textbf{48}, 9086\relax
\mciteBstWouldAddEndPuncttrue
\mciteSetBstMidEndSepPunct{\mcitedefaultmidpunct}
{\mcitedefaultendpunct}{\mcitedefaultseppunct}\relax
\EndOfBibitem
\bibitem[Laforge \emph{et~al.}(2001)Laforge, Passerone, Harris, Lambin, and
  Tosatti]{laforge}
C.~Laforge, D.~Passerone, A.~B. Harris, P.~Lambin and E.~Tosatti, \emph{Phys.
  Rev. Lett.}, 2001, \textbf{87}, 085503\relax
\mciteBstWouldAddEndPuncttrue
\mciteSetBstMidEndSepPunct{\mcitedefaultmidpunct}
{\mcitedefaultendpunct}{\mcitedefaultseppunct}\relax
\EndOfBibitem
\bibitem[Braun and Naumovets(2006)]{braun}
O.~Braun and A.~Naumovets, \emph{Surf. Sci. Rep.}, 2006, \textbf{60}, 79\relax
\mciteBstWouldAddEndPuncttrue
\mciteSetBstMidEndSepPunct{\mcitedefaultmidpunct}
{\mcitedefaultendpunct}{\mcitedefaultseppunct}\relax
\EndOfBibitem
\bibitem[Coffey and Krim(2006)]{coffey}
T.~Coffey and J.~Krim, \emph{Phys. Rev. Lett.}, 2006, \textbf{96}, 186104\relax
\mciteBstWouldAddEndPuncttrue
\mciteSetBstMidEndSepPunct{\mcitedefaultmidpunct}
{\mcitedefaultendpunct}{\mcitedefaultseppunct}\relax
\EndOfBibitem
\bibitem[Liang \emph{et~al.}(2003)Liang, Tsui, Xu, Li, and Xiao]{liang1}
Q.~Liang, O.~K.~C. Tsui, Y.~Xu, H.~Li and X.~Xiao, \emph{Phys. Rev. Lett.},
  2003, \textbf{90}, 146102\relax
\mciteBstWouldAddEndPuncttrue
\mciteSetBstMidEndSepPunct{\mcitedefaultmidpunct}
{\mcitedefaultendpunct}{\mcitedefaultseppunct}\relax
\EndOfBibitem
\bibitem[Liang \emph{et~al.}(2006)Liang, Li, Xu, and Xiao]{liang2}
Q.~Liang, H.~Li, Y.~Xu and X.~Xiao, \emph{J. Phys. Chem. B}, 2006,
  \textbf{110}, 403\relax
\mciteBstWouldAddEndPuncttrue
\mciteSetBstMidEndSepPunct{\mcitedefaultmidpunct}
{\mcitedefaultendpunct}{\mcitedefaultseppunct}\relax
\EndOfBibitem
\bibitem[Benassi \emph{et~al.}(2010)Benassi, Vanossi, Santoro, and
  Tosatti]{benassithermo}
A.~Benassi, A.~Vanossi, G.~Santoro and E.~Tosatti, \emph{Phys. Rev. B}, 2010,
  \textbf{82}, 081401(R)\relax
\mciteBstWouldAddEndPuncttrue
\mciteSetBstMidEndSepPunct{\mcitedefaultmidpunct}
{\mcitedefaultendpunct}{\mcitedefaultseppunct}\relax
\EndOfBibitem
\bibitem[Benassi \emph{et~al.}(2012)Benassi, Vanossi, Santoro, and
  Tosatti]{benassithermo2}
A.~Benassi, A.~Vanossi, G.~Santoro and E.~Tosatti, \emph{Tribol. Lett}, 2012,
  \textbf{48}, 41\relax
\mciteBstWouldAddEndPuncttrue
\mciteSetBstMidEndSepPunct{\mcitedefaultmidpunct}
{\mcitedefaultendpunct}{\mcitedefaultseppunct}\relax
\EndOfBibitem
\bibitem[Dienwiebel \emph{et~al.}(2004)Dienwiebel, Verhoeven, Pradeep, Frenken,
  Heimberg, and Zandbergen]{dienwiebel}
M.~Dienwiebel, G.~Verhoeven, N.~Pradeep, J.~Frenken, J.~Heimberg and
  H.~Zandbergen, \emph{Phys. Rev. Lett.}, 2004, \textbf{92}, 126101\relax
\mciteBstWouldAddEndPuncttrue
\mciteSetBstMidEndSepPunct{\mcitedefaultmidpunct}
{\mcitedefaultendpunct}{\mcitedefaultseppunct}\relax
\EndOfBibitem
\bibitem[Filippov \emph{et~al.}(2008)Filippov, Dienwiebel, Frenken, Klafter,
  and Urbakh]{tt}
A.~E. Filippov, M.~Dienwiebel, J.~W.~M. Frenken, J.~Klafter and M.~Urbakh,
  \emph{Phys. Rev. Lett.}, 2008, \textbf{100}, 046102\relax
\mciteBstWouldAddEndPuncttrue
\mciteSetBstMidEndSepPunct{\mcitedefaultmidpunct}
{\mcitedefaultendpunct}{\mcitedefaultseppunct}\relax
\EndOfBibitem
\bibitem[Plimpton(1995)]{lammps}
S.~Plimpton, \emph{J. Comp. Phys.}, 1995, \textbf{117}, 1\relax
\mciteBstWouldAddEndPuncttrue
\mciteSetBstMidEndSepPunct{\mcitedefaultmidpunct}
{\mcitedefaultendpunct}{\mcitedefaultseppunct}\relax
\EndOfBibitem
\bibitem[Hockney and Eastwood(1989)]{pppm}
R.~Hockney and J.~Eastwood, \emph{Computer Simulation Using Particles}, Adam
  Hilger, New York, USA, 1989\relax
\mciteBstWouldAddEndPuncttrue
\mciteSetBstMidEndSepPunct{\mcitedefaultmidpunct}
{\mcitedefaultendpunct}{\mcitedefaultseppunct}\relax
\EndOfBibitem
\end{mcitethebibliography}
\bibliographystyle{rsc}
%}

\end{document}